\documentclass[10pt]{article}
\usepackage{epsfig,graphicx}
\usepackage{times,ch2005}

\def\beq{\begin{equation}}
\def\eeq#1{\label{#1}\end{equation}}
\def\eeqn{\end{equation}}

\def\beqa{\begin{eqnarray}}
\def\eeqa#1{\label{#1}\end{eqnarray}}
\def\eeqan{\end{eqnarray}}

\def\Dslash{\not{\hbox{\kern-4pt $D$}}}
\def\dslash{\not{\hbox{\kern-2pt $\del$}}}

\newcommand{\tev}{\ensuremath{\mathrm{\,Te\kern -0.1em V}}\xspace}
\newcommand{\gev}{\ensuremath{\mathrm{\,Ge\kern -0.1em V}}\xspace}
\newcommand{\mev}{\ensuremath{\mathrm{\,Me\kern -0.1em V}}\xspace}
\newcommand{\kev}{\ensuremath{\mathrm{\,ke\kern -0.1em V}}\xspace}
\newcommand{\ev}{\ensuremath{\mathrm{\,e\kern -0.1em V}}\xspace}
\newcommand{\gevc}{\ensuremath{{\mathrm{\,Ge\kern -0.1em V\!/}c}}\xspace}
\newcommand{\mevc}{\ensuremath{{\mathrm{\,Me\kern -0.1em V\!/}c}}\xspace}
\newcommand{\gevcc}{\ensuremath{{\mathrm{\,Ge\kern -0.1em V\!/}c^2}}\xspace}
\newcommand{\mevcc}{\ensuremath{{\mathrm{\,Me\kern -0.1em V\!/}c^2}}\xspace}

\def\mus  {\ensuremath{\rm \,\mus}\xspace}

\def\mus        {\ensuremath{\,\mu{\rm s}}\xspace}    

\def\laeq{\raise.2ex\hbox{$<$}\kern-.75em\lower.9ex\hbox{$\sim$}\,}
\def\gaeq{\raise.2ex\hbox{$>$}\kern-.75em\lower.9ex\hbox{$\sim$}\,}

\begin{document}


\Title{Ground-based Gamma-Ray Observations of Pulsars and their Nebulae: Towards a New Order}
\bigskip


%
\label{deJagerStart}

%
\author{Ocker C. de Jager and Christo Venter\index{de Jager, O.C.}}
%
\address{Unit for Space Physics\\
North-West University, Potchefstroom Campus\\ 
Private Bag X6001, Potchefstroom, 2520, South Africa
}

\makeauthor\abstracts{
The excellent sensitivity and high resolution capability
of wide FoV ground-based imaging atmospheric Cerenkov telescopes allow us for the
first time to resolve the morphological structures of pulsar wind nebulae (PWN) which are older
and more extended than the Crab Nebula. VHE $\gamma$-ray observations of such extended nebulae
(with field strengths below $\sim 20\,\mu$G) probe the electron component corresponding to
the unseen extreme ultraviolet (EUV) synchrotron component, which measures electron injection from earlier evolutionary epochs. 
VHE observations of PWN therefore introduce a new window on PWN research. This review paper\footnote{To appear in the proceedings of Cherenkov 2005, Palaiseau, France} also identifies conditions for maximal VHE visbility of PWN. Regarding pulsar pulsed emission, it is becoming clear that the threshold energies of current telescopes are not sufficient to probe the pulsed $\gamma$-ray component from canonical pulsars. Theoretical estimates of pulsed $\gamma$-ray emission from millisecond pulsars (MSPs) seem to converge and it becomes clear that such detections with current 3$^{\rm rd}$ generation telescopes will not be possible, unless the geometry is favourable.
}

\section{The definition of pulsar wind nebulae (PWN)}
The term ``pulsar wind nebulae'' (PWN, or ``plerions'' - a term coined by Weiler \& Panagia \cite{wp78}) signifies a  

{\bf (1)} filled centre or blob-like form;
 
{\bf (2)} a flat radio spectrum of the form $\alpha\sim 0$ to -0.3;

{\bf (3)} a well-organised internal magnetic field; and

{\bf (4)} a high integrated linear polarisation at high radio frequencies.

Since this original classification, significant progress with PWN research has been made, especially in X-rays, so that we can add the following list of typical characteristics:

{\bf (5)} A torus and jet near the pulsar, with the direction of the jet reflecting the 
direction of the pulsar spin axis and the torus showing an underluminous region
inside a characteristic scale radius $r_s\sim 10^{17}$cm to $\sim 10^{18}$cm, 
believed to be the pulsar wind shock radius \cite{NR04};

{\bf (6)} Evidence for reacceleration of particles somewhere between the pulsar light cylinder
and $r_s$, leading to a hard X-ray spectrum with a photon index $\sim 1.5$ to 2.0 near $r_s$
(for example: $\sim 1.8$ for Crab -- \cite{mori_2004}, 1.50 for Vela -- \cite{mangano} and 
$\sim 2.0$ for 3C\,58 -- \cite{slane_2004});

{\bf (7)} Evidence for synchrotron cooling (spectral steepening) at $r>r_s$, with the
size of the PWN decreasing towards increasing energies, as seen from the Crab and several
other PWN. The photon indices of the cooled spectra range between 2.0 and 2.5;

{\bf (8)} Evidence for a correlation between the spindown power and the average spectral 
indices of the pulsar and its PWN \cite{gotthelf_2003}.

It is now important to understand how the general PWN properties relate to their
VHE $\gamma$-ray properties. We will highlight a few aspects, following a brief
discussion of the Crab Nebula:  

The Crab Nebula is considered to be the most important prototype PWN,
serving as a calibration source for hard X-rays to TeV $\gamma$-rays
and was also the first source to be seen at TeV energies \cite{weekes}.
The emission of this source is understood to be due to synchrotron-self
Compton (SSC), with multiwavelength modelling by \cite{dh92} and \cite{aa96}.
The problem with Crab-like PWN from a ground-based perspective 
is that the strong magnetic field strength in such young 
PWN results in rapid cooling, so that almost all energy is radiated as synchrotron
emission, leaving a relatively weak SSC component \cite{aa96}. 
The only reason why Crab is still bright at TeV energies, is because
the spindown flux $\dot{E}/d^2\sim 10^{38}$ ergs/s/kpc$^2$ is unusually large compared
to all other known pulsars. The X-ray photons serve as markers for VHE $\gamma$-ray emitting electrons.
For the lower $\dot{E}$ PWN, the field is much lower so that the electrons radiating extreme ultraviolet (EUV) synchrotron photons also inverse Compton (IC) scatter CMBR into the VHE $\gamma$-ray range. 

\section{Optimal young VHE $\gamma$-ray PWN}
Rapid SNR/PWN expansion within the first 1,000 years would
be a key condition to reduce the magnetic field strength fast enough so 
that synchrotron losses are minimised during the early phase of injection
while the spindown power is still quite high. If $\tau_0=P_0/2\dot{P}_0$
is the characteristic age at birth (with $P_0$ the birth period and
$\dot{P}_0$ the corresponding period derivative), the spindown power at present 
(for a pulsar braking index of $n=3$) would reduce with time as
$$
\dot{E}=\dot{E}_0\frac{1}{(1+t/\tau_0)^2}.\eqno (1)
$$
The age of the pulsar is then evaluated in terms of the ratio $t/\tau_0$:
If the pulsar is still relatively young (i.e. $t\sim \tau_0$),
then $\dot{E}$ is not much less than $\dot{E}_0$, whereas the
spindown power for old pulsars ($t\gg \tau_0$) decreases with time as $(t/\tau_0)^2$.
The evolution of the $\gamma$-ray flux with time is then evaluated
by considering a combination of the time evolution of $\dot{E}$,
and the synchroton lifetime of VHE $\gamma$-ray emitting electrons:
If $E_{\rm TeV}$ is the energy of a $\gamma$-ray (in units of TeV) as a result of 
IC scattering on the CMBR, its synchrotron lifetime in a field of strength
$B_{-5}$ (in units of $10^{-5}$ G) would be
$$
T_{\gamma}=4.8\,({\rm kyr})B_{-5}^{-2}E_{\rm TeV}^{-1/2}.\eqno (2)
$$
However, the synchrotron lifetime of an electron emitting keV
synchrotron photons would be correspondingly shorter:
$$
T_{\rm X}=1.2\,({\rm kyr})B_{-5}^{-3/2}E_{\rm KeV}^{-1/2}.\eqno (3)
$$
In regions of our galaxy where the energy density of dust with temperature $\sim 25\,$K
exceed the CMBR density significantly, lower energy electrons would
produce the same TeV $\gamma$-ray energies, so that the synchrotron
lifetime would increase relative to the number given in Eqn.~(2). 
Eqn.~(2) however breaks down in the extreme Klein-Nishina limit.
  
Clearly, if the PWN expanded fast enough up to the present time $t$, with the nebular field 
strength small enough so that VHE emitting electrons can survive the early phases
of expansion (i.e. $T_{\gamma}>t$), we should then have an optimal VHE emitter.
In this case we can tap most of the energy (in electrons) ejected by the spinning dipole.
The SSC process would then be inactive (because of the relatively low synchrotron
brightness), resulting in IC scattering on the CMBR and possibly far-infrared photons from galactic
dust grains as the dominant VHE $\gamma$-ray production mechanism. 

Gaensler et al. \cite{chandra_g09} found evidence for rapid expansion in
G\,0.9+0.1 and added this source to the PWN associated with PSR~B\,1509-58,
forming a class of rapidly expanding PWN. These were also the next two
PWN which were detected at VHE energies \cite{hess_g09,hess_1509}, and
the former detection was made despite its large distance to the galactic centre.
Du Plessis et al. \cite{duplessis} was however the first to predict that rapidly expanding young PWN (using PSR~B\,1509-58 as the prototype) should be relatively bright at very high $\gamma$-ray energies: The latter authors calculated the field strength ($\sim 8\,\mu$G) associated with such rapid expansion in the PWN of PSR~B\,1509-58 and predicted a flux at 1~TeV, which is relatively close to the recently detected 1~TeV flux, although the spectral shape is wrong as a result of the poorly constrained X-ray data available at that time.  

A recent review on young PWN with outstanding problems is given by
Chevalier \cite{chevalier_2004}. Of interest in his discussion is the addition of 3C\,58:
For both the PWN of PSR~B\,1509-58 and 3C\,58, the total energy in 
particles is expected to exceed that of the magnetic field. Thus, the
magnetisation parameter $\sigma_{\rm tot}$ (defined as the ratio of 
magnetic energy density to particle energy density) is much less than unity,
in which case IC radiation competes more favourably with synchrotron emission.
Note that the Kennel \& Coroniti \cite{KC84} model does predict a particle dominated
wind at the pulsar wind shock radius for Crab ($\sigma_s\sim 0.003$ at $r_s\sim 0.1$ pc),
but equiparition is reached further downstream near the bright part of the torus,
which is not conducive towards the survival of VHE emitting electrons.

The ideal VHE PWN emitters therefore share at least one or more of the following
conditions:

{\bf (a)} The overall (total) wind magnetization parameter of the PWN $\sigma_{\rm tot}$ should be  much less
than unity (i.e. a particle dominated wind);

{\bf (b)} Rapid expansion of the PWN during its early phases of high power input from the
pulsar results in the survival of the majority of VHE emitting electrons 
since early epochs. This condition, and the former, are formally consistent with
synchrotron losses being much less important relative to IC;

{\bf (c)} The ideal condition (which includes the first two conditions) 
is to have the lifetime of VHE radiating particles comparable to, or longer
than, the age of the system ($T_{\gamma}\sim t$), surviving even the earlier epochs
when the field was stronger, so that the total amount of energy in electrons in the
PWN is a significant fraction of the maximal rotational kinetic energy of the neutron star $I\dot{\Omega}_0^2/2$ at birth. Only adiabatic
losses are then the main source of losses. In this case we do not expect to
see an energy dependence of the PWN size with changing $\gamma$-ray energy --
a well known phenomenon for PWN where the lifetime of particles exceed the
age of the system.   

{\bf (d)} The spindown power at birth should be much larger than the present
spindown power, so that, assuming the abovementioned conditions hold,
relic VHE emitting electrons stored in the reservoir since birth may
still contribute to the present VHE $\gamma$-ray flux. In this case
we see the integrated spindown power over a time scale $T_{\gamma}$ into the past,
which collects electrons from the epoch of much higher spindown power.

\section{PWN visibility: the `N-zone' empirical approach}  
A simplified approach to estimate the $\gamma$-ray visibility of PWN
is to take the total X-ray flux of the PWN, invert the X-ray spectrum to
get the electron spectrum, while assuming a single constant 
field strength. The IC flux 
(as a result of scattering on the CMBR and diffuse galactic photon fields)
is then calculated. This phenomological approach
is probably too simplistic for PWN showing significant gradients in the 
magnetic energy density across the emission volume.

De Jager \cite{d05} introduced the following
quasi-phenomological model-based estimate of the expected $\gamma$-ray flux
from G\,0.9+0.1, using what we define here as the
`N-zone' model approach: Assume the X-ray spectral index $\Gamma_i$
and flux normalisation $F_i$ are known for $N$ radial shells (determined from
X-ray observations), corresponding to radii $R_i$, $i=1, \ldots, N$.
Since the confining pressure eventually tends to decelerate the flow, the 
flow speed $V_i$ should reduce with increasing radius. 
The total number of electrons in the last (`N-th') shell should 
then also be large compared to the total, since the 
electron residence time, $\propto 1/V_N$, is maximal for the last shell, 
leading to this outer volume (between radii $R_{N-1}$ and $R_N$)
contributing mostly to the VHE emission via
IC scattering. Furthermore, even though we are dealing with an
unknown magnetic field strength profile with $R_i$, we may assume that
the field strength does not change inside the last shell, so that we 
may apply the constant field strength (`one-zone model') approach
to the bulk of PWN electrons, resident in the last shell.
This should lead to a conservative estimate of the VHE luminosity.
One can then either use the full numerical method (using the full cross section for IC scattering)
or the `pocket calculator' approximations of \cite{d05}
to `invert' the X-ray spectrum of the `N-th' zone to give the approximate IC spectrum 
corresponding to this important VHE emitting volume.  

However, care should be taken with this procedure: do the electrons contributing to the 
X-ray domain also contribute to the VHE domain?
To identify the same population of electrons which produces synchrotron and IC $\gamma$-rays
via scattering on the CMBR, we can use the expression rederived by \cite{aak}, 
giving the synchrotron photon energy in terms of the IC scattered $\gamma$-ray energy $E_{\rm TeV}$
in units of TeV for a field strength of $B_{-5}$ in units of $10\,\mu$G:
$$
E_{\rm syn}=70B_{-5}E_{\rm TeV}\,{\rm eV}.\eqno(4)
$$
It is thus clear that VHE emission from PWN with these typical field strengths
correspond to synchrotron emission in the Extreme Ultraviolet to soft X-ray domain.
For the IC scattering of 25\,K dust photons in regions of the galaxy where 
this component is important \cite{d95,duplessis,d05}, the constant will be nearly ten
times smaller, but extreme Klein-Nishina corrections will have to be added.

Wide FoV optical observations of PWN (if possible) are also essential to 
complement VHE observations, since the electrons associated with the optical
have even longer lifetimes than the VHE $\gamma$-ray emitting population
and probe the time-integrated history even closer to the birth of the PWN. GLAST 
observations of PWN in the 10 to 100~GeV domain should also be
helpful to provide complementary information.

Finally, one should be careful when extrapolating X-ray spectra into the optical domain
(even for the `N-th' shell) since evolutionary effects may
create spectral states, which are quite different from our extrapolations.

\section{G\,0.9+0.1 and 3C\,58: A comparative study.} 

HESS recently identified the galactic centre region composite SNR G\,0.9+0.1 as one of the faintest
VHE $\gamma$-ray sources \cite{hess_g09}. For this unresolved VHE source 
(diameter: 2 arcmin in radio and X-rays), 
\cite{d05} defined the outer `region 3' of the {\it XMM-Newton} source \cite{g09_xmm} 
as the `N-th' zone and found that a field strength between 10 and $14\,\mu$G
can explain the VHE spectrum if the X-ray spectrum ($\Gamma_X\sim 2.3$)
extrapolates unbroken into the EUV range, so that we observe a similar photon 
index in $\gamma$-rays.
 
Whereas `region 1' contains the much harder uncooled spectral component (close
to the injection region at $r_s$), `region 3' in X-rays and the HESS spectrum are consistent with
the synchrotron-cooled equivalent. The fact that we do not see the cooling
break in the HESS range should place important constraints on the evolutionary history
of G\,0.9+0.1. The reader is also referred to \cite{d05} for a discussion on the lack of a
spectral break in the HESS spectrum \cite{hess_1509} of the PWN in PSR~B\,1509-58.

The northern hemisphere 3C\,58 is an important member of this class of PWN: 
whereas it shares `Crab-like' features, its extended nature should place it
in the category of G\,0.9+0.1 and the PWN of PSR~B\,1509-58, making it an
important object for VHE $\gamma$-ray studies. It is probably related to the 
Medieval SNR of 1181\,AD, resulting in \cite{murray} deriving an initial spin 
period of $\sim 60$ ms, given its present period of 65 ms. 

Comparing 3C\,58 with G\,0.9+0.1 yields the following:
 
(a) The inferred spindown power of G\,0.9+0.1 ($\sim 1.5\times 10^{37}$ ergs/s -- 
\cite{sid00}) is comparable to that of 3C\,58 ($2.7\times 10^{37}$ ergs/s -- \cite{murray}); 
(b) The 2-10 keV X-ray luminosities are $5.4\times 10^{34}d_{8.5}^2$ ergs/s 
for G\,0.9+0.1 \cite{g09_xmm} and $\sim 1.3\times 10^{34}d_{3.2}^2$ ergs/s 
(derived from \cite{xmm_3c58}) for 3C\,58; 
(c) The dimensions of 3C\,58 are $10\times 6$ pc$^2$, which is larger compared to 
G\,0.9+0.1, which has dimensions of $\sim 5\times 5$ pc$^2$;
(d) both show evidence of synchrotron cooling,
with the X-ray photon index increasing towards the edge of the PWN; 
(e) both show an approximate constant X-ray flux in equal radial intervals $dR$,
which would hint at similar radial magnetic field strength profiles; 
(f) for both we find that the size of the PWN at the softest X-ray energies
is comparable to the size of the radio PWN, which would indicate a cooling
timescale comparable to, or, longer than the age of the PWN/SNR. In the case
of the Crab Nebula, this condition only holds for the radio to far infrared
emitting electrons, which do not radiate VHE $\gamma$-rays.

Inspecting Figure 2 of \cite{xmm_3c58} for 3C\,58 shows that
the X-ray and radio sizes are equal for photon energies below
1 keV, whereas the X-ray size shrinks with increasing energy above1 keV.
This means that electrons radiating 1 keV synchrotron photons at the radio boundary
should have electron lifetimes equal to the 820 year age of the SNR, and 
from Eqn.~(3) we see that this corresponds to a field strength of 
$B_N\sim 13\,\mu$G, which is comparable to the field strengths derived 
for G\,0.9+0.1 and PSR~B\,1509-58 from HESS observations 
\cite{hess_g09,hess_1509,d05}. The X-ray flux of 3C\,58 is twice that
of G\,0.9+0.1, but the major uncertainty in making VHE $\gamma$-ray 
flux predictions is the spectral state of the VHE emitting electrons,
which would correspond to the unseen EUV synchrotron component.
One point of concern is the fact that the spindown power did not change
significantly since birth (assuming the historical association with
the Medieval SNR is correct), in which case it was not possible to
collect significant amounts of relic electrons from an earlier epoch 
of much larger spindown power, despite the favourable condition of rapid expansion and
negligible synchrotron losses.
Thus, whereas conditions {\bf (a), (b)} and {\bf (c)} appears to be ideal
for 3C\,58, condition {\bf (d)} may prove to be problematic from a detection
perspective. 

\section{Middle-aged PWN: new progress}
Vela X, the PWN of the Vela SNR, was the catalyst for the concept of PWN
evolution \cite{WP80,RC84}, as well as the interpretation of middle-aged PWN
that are offset from their parent pulsars \cite{b01}: The basic idea is that
a SNR shell usually expands into an inhomogeneous ISM, resulting in an asymmetric
reverse shock crushing back into the PWN. The reverse shock originating from the 
section where the ISM pressure was largest, will crush first into the PWN, 
creating an offset nebula, as seen from Vela X \cite{b01}. If the electron lifetime
is longer than the crushing timescale, we can still see relic electrons radiating
for relatively long times in a lower B environment. The radio
and VHE $\gamma$-ray emitting electrons will create the appearance of extended shapes
that are larger than the X-ray shapes, unless the X-ray emitting electrons also
have relatively long lifetimes. 

ROSAT and XMM observations of PSR~B\,1823-13 resulted in the detection
of the PWN G\,18.0-0.7, which was found to be offset by a few arcmin south of its
parent pulsar. This prompted Gaensler et al. \cite{xmm_1823} to  
invoke the offset explanation for Vela X (as proposed by \cite{b01}) as applicable to G\,18.0-0.7.
This association was further boosted by HESS observations of this region, showing the detection
of highly significant \textit{resolved} VHE $\gamma$-ray emission, which is also displaced south of the pulsar 
\cite{hess_1825}. Whereas background noise in the XMM image makes it difficult to determine 
the exact X-ray size ($\gaeq 5$ arcmin), the VHE $\gamma$-ray source (HESS\,J1825-137) 
size is clearly larger (about 0.5 degrees.) The symmetric compact nebular spectrum 
with a photon index of $\sim 1.6$ eventually cools towards the asymmetric offset nebula, with 
photon index $\sim 2.3$. The latter is also consistent with the VHE $\gamma$-ray photon index 
of $\sim 2.4$ (interpreted as IC scattering on the CMBR), which hints at the importance
of evolutionary effects.

A systematic investigation of middle-aged pulsars at the edges of resolved 
\textit{centre-filled} VHE $\gamma$-ray sources, combined with follow-up radio and 
X-ray imaging information, should result in the identification of a new class of 
PWN, with Ground-Based Gamma-Ray Astronomy taking
the lead in this new direction. GLAST operations at its highest energies (where the
angular resolution is best) is also expected to make a contribution to this field.

\section{Pulsars: pulsed VHE $\gamma$-ray emission}

The possibility of detecting canonical $\gamma$-ray pulsars with present 3$^{\rm rd}$ generation ground-based $\gamma$-ray telescopes was discussed some time ago (see e.g.\ \cite{deJager01}), but the fact that the spectral cutoffs are below the thresholds of the current generation of telescopes forces us to wait for GLAST and 4$^{\rm th}$ generation ground-based telescopes (hopefully operating at thresholds below 30 GeV) to study the physics associated with spectral cutoffs in high-B (canonical) pulsars.

Studies regarding visibility of VHE $\gamma$-rays from millisecond pulsars (MSPs) were boosted by the pioneering work of Usov \cite{Usov83} who showed that $\gamma$-rays with energies of $\sim$ 100~GeV may be expected to escape from the magnetosphere of an MSP. A number of studies furthermore centre on $\gamma$-ray MPS visibility, e.g.\ \cite{Rudak99,Bulik00}, and specifically within a GR framework, e.g.\ \cite{Harding-ch2005,Harding05,Venter05,Frackowiak05} (See Table~\ref{tab:deJager-tab1} for a comparison of different authors' converging results for curvature radiation (CR) cutoff energies). 

Two factors determine the visibility of pulsars: spectral cutoff energy and observable flux. If the cutoff energy is \gaeq the threshold energy of the telescope, and the flux sensitivity is \laeq 0.05$\dot{E}/d^2$, with $d$ the distance, VHE $\gamma$-rays from a pulsar should be visible. One of the main uncertainties when modelling pulsars, however, is the geometry (i.e.\ magnetic inclination angle $\chi$ between the magnetic and spin axes, and the angle $\zeta$ of the observer's line of sight with respect to the spin axis) of the pulsar, which significantly influences predictions of cutoff energy. 

Historically it therefore appeared as if MSPs would provide suitable laboratories for current (then future) 3$^{\rm rd}$ generation telescopes. However, new calculations and final calibration of 3$^{\rm rd}$ generation telescopes indicate that  4$^{\rm th}$ generation Cerenkov telescopes (see e.g.\ \cite{Konopelko05}) and GLAST \cite{Thompson04} are generally needed in order to observe $\gamma$-ray MSPs, due to the relatively low cutoff energies \cite{Harding-ch2005,Harding05,Bulik99}. 

\begin{table}[h]
  \begin{tabular}{|c||c|c|c||c|c|c|}
  \hline
   Reference & \multicolumn{3}{|c||}{PSR~J\,0437-4715} & \multicolumn{3}{c|}{PSR~B\,1821-24} \\
   \hline
   \, & $\chi\,(^\circ)$ & $\epsilon^{\max}_{\rm CR}$~(GeV) & Screened & $\chi\,(^\circ)$ & $\epsilon^{\max}_{\rm CR}$~(GeV) & Screened\\
   \hline
   \hline
   \cite{Venter05} & all & \laeq 17 & no & \, & \, &\,\\
   \hline
   \cite{Venter05} & 20; 35 & $\sim 1-10$ & no & \, & \, &\,\\
   \hline
   \cite{Venter05b} & all & $\sim 1-20$ & no &\, & \, &\,\\
   \hline
   \cite{Venter05b} & 10 & \laeq 10 & no &\, & \, &\,\\
   \hline
   \cite{Harding05} & 10 & $\sim 5$ & no & 50 &  $\sim 43$ & yes\\
   \hline
   \cite{Frackowiak05} & 20; 35 & $\sim 10$ & no & \, & \, &\, \\
   \hline
   \cite{Frackowiak05b} & 35 & $\sim 10$ & no & 50 &  \gaeq~100 & no\\
   \hline
   this paper & 10; 20 & \laeq 10 & no & 50 &  \laeq~70 & yes\\
   \hline
   this paper & 35 & \laeq 1 & no & all & $\sim 0.1-150$ & yes\\
   \hline   
  \end{tabular}
  \caption{Comparing different authors' converging results for peak CR energy, for similar pulsar parameters.}\label{tab:deJager-tab1}
\end{table}

\subsection{The unscreened case}
The results in this paragraph build on those described in \cite{Venter05,Venter05b}. (See e.g.\ \cite{Harding05,Frackowiak05,Frackowiak05b} for similar studies). We use the general relativistic B- and E-fields in the frame corotating with the pulsar \cite{Muslimov92,Muslimov97,Harding98,Harding02} and only consider the dominant CR component of $\gamma$-radiation.
\begin{figure}[htb]
  \begin{center}
   \epsfig{file=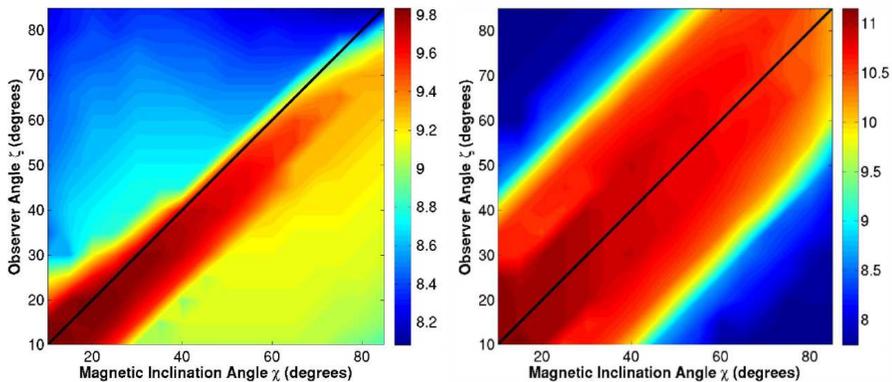,width=120mm}
    \caption{Contour plot of $\log_{10}$ of the CR cutoff energy in units of eV vs.\ observer angle $\zeta$ and magnetic inclination angle $\chi$. The left hand side is for PSR~J\,0437-4715, representing the ``unscreened case'', while the right hand side is for PSR~B\,1821-24, representing the ``screened case''. We used pulsar radius \mbox{$R = 10^6$ cm}, moment of inertia $I = 3\times10^{45}$ g.cm$^2$, and mass $M = 1.58M_\odot$.}\label{fig:deJager-fig1}
  \end{center}
\end{figure}
 
In \cite{Harding02} a critical pulsar spindown power $\dot{E}_{\rm break}$ is given. For pulsars with $\dot{E} = I\Omega\dot{\Omega} < \dot{E}_{\rm break}$ (with $\Omega = 2\pi/P$, $P$ the period, and $I$ the moment of inertia), no screening due to magnetic pair production will take place above the polar cap (PC) (within the framework of the Harding-Muslimov PC-type models). We modelled the MSP PSR~J\,0437-4715 and used the ``unscreened'' version of the E-field applicable to this case. 

Because this model takes the observer and magnetic inclination angles into account, it is possible to find the CR cutoff energy as a function of $\chi$ and $\zeta$. The result of this calculation for PSR~J\,0437-4715 is shown in the left panel of Figure~\ref{fig:deJager-fig1}. The CR cutoff energies were estimated from $E^2(dN/dE)$-spectra which we calculated for different combinations of $\chi$ and $\zeta$ in the range [10,85].

Maximum CR cutoff energies \laeq 10~GeV (for the parameters listed in the caption of figure~\ref{fig:deJager-fig1}) are obtained when $\chi \sim \zeta$. This should be expected, since the observer sweeping close to the magnetic axis will sample radiation due to particle acceleration by the highest values of the electric potential. Equality does not hold exactly, as these maxima are below the $y = x$ line. This is due to the asymmetric character of the E-field with respect to the magnetic azimuthal angle $\phi$. The energy maxima occuring at small values of $\chi$ also agrees with the fact that the E-field has a  $\cos^2\chi$-term dominating at low altitudes, with a $\sin^2\chi$-term coming into play at larger altitudes.

\subsection{The screened case}
We chose PSR~B\,1821-24 as an example of an MSP with a screened electric potential. We used the approximation of \cite{Dyks00} and chose the height of the pair formation front (PFF) $h = R_{PC} \sim (\Omega R^3/c)^{1/2}$. 

The CR cutoff energy \laeq 150~GeV as a function of $\chi$ and $\zeta$ for PSR~B\,1821-24 is also shown in Figure~\ref{fig:deJager-fig1} (right panel). This case seems to be more symmetric with respect to the $y = x$ line and reaches the highest energy values at small $\chi$ due to the fact that the E-field $\propto \cos\chi$.

What is very important from Figure 1 (for both the ``screened'' and ``unscreened'' cases), is that a MSP is not expected to be visible for current 3$^{\rm rd}$ generation telescopes
(within the GR field definition of \cite{Muslimov92,Muslimov97,Harding98,Harding02}),
unless the observer sweeps through the magnetic axis, while having the angle between this
magnetic axis and spin axis substantially less than $\sim 45$ degrees.

Finally, the detection of ``pair-starved'' (low spindown) pulsars is very important to probe the naked acceleration potential without the complicating screening effects due to pair production. It is however clear that we will have to wait for GLAST and 4$^{\rm th}$ generation telescopes to answer some of the most fundamental questions related to rotating astrophysical dynamos.

\section*{Acknowledgments}
This publication is based upon work supported by the South African National Research Foundation under
Grant number 2053475.

\bibliographystyle{h-elsevier3}
\bibliography{Paris_2}

\begin{thebibliography}{10}

\bibitem{wp78}
K.W. {Weiler} and N. {Panagia},
\newblock A \& A 70 (1978) 419.

\bibitem{NR04}
C.Y. {Ng} and R.W. {Romani},
\newblock ApJ 601 (2004) 479.

\bibitem{mori_2004}
K. {Mori} et~al.,
\newblock ApJ 609 (2004) 186.

\bibitem{mangano}
V. {Mangano} et~al.,
\newblock A \& A 436 (2005) 917.

\bibitem{slane_2004}
P. {Slane} et~al.,
\newblock ApJ 616 (2004) 403.

\bibitem{gotthelf_2003}
E.V. {Gotthelf},
\newblock ApJ 591 (2003) 361.

\bibitem{weekes}
T.C. {Weekes} et~al.,
\newblock ApJ 342 (1989) 379.

\bibitem{dh92}
O.C. {de Jager} and A.K. {Harding},
\newblock ApJ 396 (1992) 161.

\bibitem{aa96}
A.M. {Atoyan} and F.A. {Aharonian},
\newblock A \& A Suppl. 120 (1996) C453.

\bibitem{chandra_g09}
B.M. {Gaensler}, M.J. {Pivovaroff} and G.P. {Garmire},
\newblock ApJL 556 (2001) L107.

\bibitem{hess_g09}
F. {Aharonian [HESS Collaboration]},
\newblock A \& A 432 (2005) L25.

\bibitem{hess_1509}
F. {Aharonian [HESS Collaboration]},
\newblock A \& A 435 (2005) L17.

\bibitem{duplessis}
I. {Du Plessis, et al.},
\newblock ApJ 453 (1995) 746.

\bibitem{chevalier_2004}
R.A. {Chevalier},
\newblock Adv. Space Res. 33 (2004) 456.

\bibitem{KC84}
C.F. {Kennel} and F.V. {Coroniti},
\newblock ApJ 283 (1984) 710.

\bibitem{d05}
O.C. {de Jager},
\newblock AIP Conf. Proc. 801: Astrophysical Sources of High Energy Particles
  and Radiation, p. 298, 2005.

\bibitem{aak}
F.A. {Aharonian}, A.M. {Atoyan} and T. {Kifune},
\newblock MNRAS 291 (1997) 162.

\bibitem{d95}
O.C. {de Jager, et al.},
\newblock Proc. 24th Int. Cosmic Ray Conf. 4 (1995) 528.

\bibitem{g09_xmm}
D. {Porquet}, A. {Decourchelle} and R.S. {Warwick},
\newblock A \& A 401 (2003) 197.

\bibitem{murray}
S.S. {Murray} et~al.,
\newblock ApJ 568 (2002) 226.

\bibitem{sid00}
L. {Sidoli, et al.},
\newblock A \& A 361 (2000) 719.

\bibitem{xmm_3c58}
F. {Bocchino} et~al.,
\newblock A \& A 369 (2001) 1078.

\bibitem{WP80}
K.W. {Weiler} and N. {Panagia},
\newblock A \& A 90 (1980) 269.

\bibitem{RC84}
S.P. {Reynolds} and R.A. {Chevalier},
\newblock ApJ 278 (1984) 630.

\bibitem{b01}
J.M. {Blondin}, R.A. {Chevalier} and D.M. {Frierson},
\newblock ApJ 563 (2001) 806.

\bibitem{xmm_1823}
B.M. {Gaensler} et~al.,
\newblock ApJ 588 (2003) 441.

\bibitem{hess_1825}
F. {Aharonian [HESS Collaboration]},
\newblock A \& A  (2005) \textit{accepted for publication}, astro-ph/0510394.

\bibitem{deJager01}
O.C. {de Jager} et~al.,
\newblock AIP Conf.\ Series 558: High Energy Gamma-Ray Astron., p. 613, 2001.

\bibitem{Usov83}
V.V. {Usov},
\newblock Nature 305 (1983) 409.

\bibitem{Rudak99}
B. {Rudak} and J. {Dyks},
\newblock MNRAS 303 (1999) 477.

\bibitem{Bulik00}
T. {Bulik}, B. {Rudak} and J. {Dyks},
\newblock MNRAS 317 (2000) 97.

\bibitem{Harding-ch2005}
A.K. {Harding}, V.V. {Usov} and A.G. {Muslimov},
\newblock \textit{these proceedings}  (2005), astro-ph/0510135.

\bibitem{Harding05}
A.K. {Harding}, V.V. {Usov} and A.G. {Muslimov},
\newblock ApJ 622 (2005) 531.

\bibitem{Venter05}
C. {Venter} and O.C. {de Jager},
\newblock ApJL 619 (2005) L167.

\bibitem{Frackowiak05}
M. {Fr{\c a}ckowiak} and B. {Rudak},
\newblock Adv. Space Res. 35 (2005) 1152.

\bibitem{Konopelko05}
A. {Konopelko},
\newblock Astropart. Phys. 24 (2005) 191.

\bibitem{Thompson04}
D.J. {Thompson},
\newblock New Astron. Rev. 48 (2004) 543.

\bibitem{Bulik99}
T. {Bulik} and B. {Rudak},
\newblock Astrophys. Lett. Comm. 38 (1999) 37.

\bibitem{Venter05b}
C. {Venter}, O.C. {de Jager} and A. {Tiplady},
\newblock AIP Conf. Proc. 801: Astrophysical Sources of High Energy Particles
  and Radiation, p. 278, 2005.

\bibitem{Frackowiak05b}
M. {Fr{\c a}ckowiak} and B. {Rudak},
\newblock Memorie della Societa Astronomica Italiana 76 (2005) 523.

\bibitem{Muslimov92}
A.G. {Muslimov} and A.I. {Tsygan},
\newblock MNRAS 255 (1992) 61.

\bibitem{Muslimov97}
A.G. {Muslimov} and A.K. {Harding},
\newblock ApJ 485 (1997) 735.

\bibitem{Harding98}
A.K. {Harding} and A.G. {Muslimov},
\newblock ApJ 508 (1998) 328.

\bibitem{Harding02}
A.K. {Harding}, A.G. {Muslimov} and B. {Zhang},
\newblock ApJ 576 (2002) 366.

\bibitem{Dyks00}
J. {Dyks} and B. {Rudak},
\newblock A \& A 362 (2000) 1004.

\end{thebibliography}

\label{deJagerEnd}
\end{document}